\documentclass[conference]{IEEEtran}
\IEEEoverridecommandlockouts

\usepackage{cite}
\usepackage{amsmath,amssymb,amsfonts}
\usepackage{graphicx}
\usepackage{textcomp}
\usepackage{xcolor}

\usepackage{booktabs}
\usepackage{multirow}

\usepackage{algorithm}
\usepackage{algpseudocode}

\usepackage[absolute,overlay]{textpos}

\def\BibTeX{{\rm B\kern-.05em{\sc i\kern-.025em b}\kern-.08em
    T\kern-.1667em\lower.7ex\hbox{E}\kern-.125emX}}

\usepackage{background}
\usepackage{xcolor}
\usepackage{hyperref}

\backgroundsetup{
  scale=1,
  color=black,
  opacity=1,
  angle=0,
  position=current page.south,
  vshift=10pt,
  contents={\textcolor{red}{© 2026. For personal use only. Published version under doi:  \href{https://doi.org/10.1109/ISNCC70543.2026.11693877}{10.1109/ISNCC70543.2026.11693877}. }}
}

\begin{document}

\title{Knowledge Distillation for Intelligent Softwarized Networks: Advances and Open Challenges}

\author{\IEEEauthorblockN{Mohamed Ali Zormati}
\IEEEauthorblockA{\textit{Université de Technologie de Compiègne} \\
\textit{CNRS, Heudiasyc UMR 7253}\\
Compiègne, France \\
mohamed-ali.zormati@hds.utc.fr}
\and
\IEEEauthorblockN{Ghada Jaber}
\IEEEauthorblockA{\textit{Université de Technologie de Compiègne} \\
\textit{CNRS, Heudiasyc UMR 7253}\\
Compiègne, France \\
ghada.jaber@hds.utc.fr}
\and
\IEEEauthorblockN{Hicham Lakhlef}
\IEEEauthorblockA{\textit{Université de Bordeaux, CNRS} \\
\textit{Bordeaux INP, LaBRI UMR 5800}\\
Talence, France \\
hicham.lakhlef@labri.fr}
}

\maketitle

\begin{abstract}
The increasing adoption of software defined networking and network function virtualization, combined with rapid advances in Machine Learning (ML), is driving the evolution toward intelligent network softwarization across cloud, edge, and distributed environments. However, deploying complex learning models in such heterogeneous environments introduces challenges in latency, computation, and energy consumption. Knowledge Distillation (KD) has emerged as a promising approach to enable lightweight and efficient intelligence by transferring knowledge from high-capacity teacher models to compact student models. Despite its extensive study in general ML domains, the integration of KD into intelligent softwarized networks remains fragmented and underexplored. In this paper, we review and classify recent efforts that incorporate KD within softwarized networks, analyze current trends and limitations, and outline open challenges and future directions toward scalable, adaptive, and energy-aware distillation mechanisms.
\end{abstract}


\section{Introduction}
The networking landscape is undergoing a fundamental architectural shift driven by the emergence of network softwarization \cite{setiawanEnergyEfficientSoftwarizedNetworks2025}, which migrates network functions from rigid hardware appliances to programmable and virtualized software components. Core softwarization technologies, such as Software-Defined Networking (SDN) and Network Function Virtualization (NFV), decouple the control and data planes, virtualize network functions, and enable the dynamic provisioning of services over generic infrastructure \cite{zormatiReviewAnalysisRecent2024}. Along with advances in other enablers, such as programmable data planes and network slicing, softwarization has redefined networks as flexible, service-oriented platforms rather than static infrastructures. This paradigm facilitates rapid service deployment, fine-grained configurability, global network visibility, and scalable management across diverse environments \cite{ammarInDepthSurveyVirtualization2024a}, spanning large-scale telecommunications systems, distributed computing infrastructures, and Internet of Things (IoT) networks.

Recently, leveraging network programmability, we have witnessed the evolution toward intelligent network softwarization \cite{zormatiReviewAnalysisRecent2024}, where Machine Learning (ML) and Deep Learning (DL) models are embedded directly into network operations \cite{liuBlockchainMachineLearning2020}. Intelligent mechanisms are increasingly adopted for tasks such as routing optimization, traffic classification, anomaly detection, and service function chain orchestration. This intelligence spans control planes, data planes, and virtualized infrastructures across cloud, edge, and distributed environments, enabling adaptive, autonomous, and context-aware network behavior. However, integrating complex learning models into softwarized environments introduces substantial computational, memory, and energy overheads \cite{ridwanApplicationsMachineLearning2021a}, which become particularly critical in latency-sensitive and resource-constrained scenarios where deployment decisions must balance accuracy, inference latency, and energy efficiency. Furthermore, from a green computing perspective \cite{setiawanEnergyEfficientSoftwarizedNetworks2025}, the deployment of computation-intensive models raises significant concerns regarding energy consumption, carbon footprint, and the long-term sustainability of network infrastructures.

To mitigate these limitations by reducing model complexity and inference latency, Knowledge Distillation (KD) has emerged as an effective model compression and knowledge transfer paradigm \cite{gouKnowledgeDistillationSurvey2021}. KD enables the transfer of knowledge from a large, high-capacity teacher model to a lightweight student model. Originally developed in general ML domains \cite{liWhenObjectDetection2023}, it has demonstrated effectiveness in reducing model size, accelerating inference, and lowering energy consumption while preserving predictive performance. These properties make KD particularly suitable for intelligent softwarized networks, where efficient and lightweight inference is critical, especially in resource-constrained environments. Beyond conventional compression, KD provides architectural flexibility by enabling hierarchical learning and distributed intelligence \cite{liContinualLearningKnowledge2025}.

Despite research efforts that integrate KD into networking contexts, the existing body of work remains fragmented. Prior works have extensively examined network softwarization paradigms \cite{setiawanEnergyEfficientSoftwarizedNetworks2025}, intelligent softwarized networks \cite{zormatiReviewAnalysisRecent2024}, and KD techniques \cite{gouKnowledgeDistillationSurvey2021} in isolation. In parallel, several recent works have incorporated KD into particular networking functions, including routing optimization \cite{ameurExploringTeacherStudentLearning2025}, network slicing \cite{kwantwiPersonalizedFederatedLearning2025}, service function chaining \cite{tangAnomalyDetectionService2024}, in-network classification \cite{xieEmpoweringInNetworkClassification2024}, and anomaly detection \cite{elormkuadeyDeepFedKDADDeepFederated2023}. However, these efforts remain confined to individual use cases and provide limited analysis of cross-layer integration and deployment trade-offs.

To the best of our knowledge, there is currently no comprehensive survey that examines KD as an enabling mechanism for intelligent network softwarization. To address this gap, we provide a structured overview of KD in intelligent softwarized networks by classifying existing works across functional layers, analyzing design choices and trade-offs, identifying limitations, and outlining open challenges and future directions.

To ensure a structured analysis, we adopt a systematic literature review approach. Our review is based on recent studies identified through major digital libraries using keywords such as “knowledge distillation”, “network softwarization”, “SDN”, “NFV”, and “edge computing”, and selected according to relevance, methodology, and evaluation context.

The remainder of this paper is organized as follows. Section II provides the background on network softwarization, intelligent network softwarization, and knowledge distillation. Section III reviews and classifies existing works. Section IV discusses open challenges and outlines future research directions. Finally, Section V concludes the paper.

\section{Background}
In this section, we introduce the fundamentals of network softwarization with an emphasis on SDN and NFV, intelligent network softwarization through ML techniques, and KD as a model compression and knowledge transfer paradigm.

\subsection{Network softwarization}
Network softwarization represents an architectural shift from hardware-centric infrastructures to programmable, flexible, and service-oriented systems \cite{setiawanEnergyEfficientSoftwarizedNetworks2025}. Its primary objective is to establish an open networking ecosystem that decouples hardware from software, enabling rapid service evolution. By implementing network functions as software components on general-purpose infrastructure, softwarization enhances service agility, expands networking capabilities, and simplifies network development and maintenance \cite{ammarInDepthSurveyVirtualization2024a}.

Several key technologies jointly enable network softwarization \cite{zormatiReviewAnalysisRecent2024}, supporting end-to-end programmability, service agility, and scalable orchestration. These include SDN, which decouples control and data planes for centralized management; NFV, which virtualizes network functions on commodity hardware; network slicing, which creates multiple logical networks over shared infrastructure; programmable data planes (e.g., P4-based switches \cite{kianpishehSurveyInNetworkComputing2023}) that enable fine-grained in-network processing; and cloud-edge integration, which supports distributed deployment of virtualized functions across centralized and edge environments.

However, network softwarization introduces challenges. The added programmability and virtualization layers increase management complexity, raise scalability concerns, and incur orchestration overhead and performance variability. Moreover, emerging use cases, such as Time Sensitive Networking (TSN) \cite{seolTimelySurveyTimeSensitive2021}, require real-time decision-making and dynamic resource optimization, capabilities that traditional approaches cannot efficiently support. These characteristics also introduce constraints on computation, latency, and resource management, particularly when integrating intelligent functions into distributed environments. These limitations have driven the integration of ML techniques into softwarized networks.

\subsection{Machine learning in softwarized networks}
The evolution of network softwarization has been accompanied by the integration of ML techniques to enable intelligent network management \cite{liuBlockchainMachineLearning2020}. This paradigm embeds data-driven algorithms into programmable networks to support self-optimization, self-configuration, and self-healing capabilities \cite{ridwanApplicationsMachineLearning2021a}. ML adoption in softwarized networks is driven by increasing complexity and heterogeneous environments. Functionally, ML techniques can be grouped into prediction, decision-making, and distributed intelligence.

Supervised and unsupervised learning techniques have been widely adopted for tasks such as traffic classification, Quality of Service (QoS) prediction, congestion detection, and fault diagnosis \cite{ahmadMachineLearningMeets2020}. These approaches primarily address prediction tasks, where the objective is to infer network states or performance metrics from observed data. With the increasing availability of large-scale operational data, DL architectures \cite{abbasiDeepLearningNetwork2021a} have further improved predictive accuracy and enabled end-to-end optimization of networking functions.

Reinforcement Learning (RL) has emerged as a particularly powerful framework for decision-making in softwarized networks \cite{ridwanApplicationsMachineLearning2021a}. By modeling network control problems as Markov decision processes, RL enables adaptive routing, dynamic spectrum allocation, and autonomous slice management through continuous interaction with the environment. These approaches are particularly suited for control tasks, where decisions must be continuously adapted to changing network conditions. The integration of Deep Reinforcement Learning (DRL) and multi-agent learning \cite{liApplicationsMultiAgentReinforcement2022} has further improved scalability in distributed and heterogeneous networks.

More recently, Generative Artificial Intelligence (GenAI) and Federated Learning (FL) have expanded the scope of intelligent softwarization. Generative models support intent-driven networking and semantic-level reasoning \cite{vuApplicationsGenerativeAI2025}, while federated learning enables distributed and privacy-preserving training \cite{limFederatedLearningMobile2020}. In particular, FL enables distributed intelligence, where learning and inference are performed collaboratively across cloud, edge, and network nodes without centralizing data. These approaches contribute to scalable, collaborative, and data-efficient intelligence in modern networks.

Despite their effectiveness, ML-driven network functions often rely on large and computation-intensive models. Deploying such models across heterogeneous environments, particularly at the network edge, introduces strict constraints on latency, computation, and energy consumption \cite{ridwanApplicationsMachineLearning2021a}. These constraints highlight the need for lightweight and resource-efficient learning mechanisms, where model design must explicitly consider deployment constraints and system-level trade-offs, motivating the adoption of model compression techniques.

\subsection{Knowledge distillation}
Knowledge Distillation (KD) is a model compression and knowledge transfer technique in which a compact student model is trained to replicate the behavior of a larger, high-capacity teacher model \cite{gouKnowledgeDistillationSurvey2021}. Originally introduced to reduce the computational footprint of deep neural networks while preserving predictive accuracy \cite{liWhenObjectDetection2023}, KD enables efficient inference without retraining complex architectures from scratch.

Conventional KD frameworks rely on response-based distillation \cite{gouKnowledgeDistillationSurvey2021}, where the student model is trained to match the teacher’s softened output probability distribution rather than only the ground-truth labels. The teacher’s logits are scaled using a temperature parameter to produce soft targets that encode inter-class similarities. By learning from these soft targets, the student captures both the final predictions and the teacher’s confidence structure across classes.

Beyond this classical formulation, numerous KD variants have been proposed to enhance knowledge transfer across heterogeneous architectures and deployment settings \cite{moslemiSurveyKnowledgeDistillation2024}. These include feature-based distillation, which transfers intermediate representations from the teacher’s hidden layers; relation-based distillation, which preserves structural dependencies among samples; online and mutual distillation, where multiple models collaboratively exchange knowledge during training; self-distillation, in which a model refines its own representations; and federated or distributed distillation, enabling knowledge sharing across decentralized entities without direct parameter exchange.

In resource-constrained environments, KD offers distinct advantages over traditional compression techniques such as pruning and quantization \cite{chengModelCompressionAcceleration2018a}. While pruning removes redundant parameters and quantization reduces numerical precision to decrease memory and computation requirements, KD transfers functional behavior from a high-capacity model into a compact architecture. As a result, student models often retain stronger predictive performance under strict latency, memory, and energy constraints. KD does not require architectural similarity between teacher and student models \cite{gouKnowledgeDistillationSurvey2021}, enabling heterogeneous and hardware-aware design.

KD enables efficient intelligence in softwarized networks through lightweight models, but its effectiveness depends not only on accuracy but also on inference latency, communication overhead, and teacher--student placement across the network.

\section{Knowledge Distillation in Intelligent Softwarized Networks}
In this section, we analyze works that integrate KD into intelligent softwarized networks. We organize the literature according to the network function in which KD is applied. We provide a comparative analysis by identifying common patterns, trade-offs, and system-level constraints.

\subsection{Functional taxonomy of KD integration}
In what follows, we classify existing works according to the functional layer of softwarized networks in which KD is integrated, encompassing data plane, service layer, AI-native, and distributed intelligence paradigms.

\subsubsection{In-data-plane distillation}~\\
\indent The emergence of programmable data planes has enabled in-network computation at wire speed through languages such as P4 \cite{kianpishehSurveyInNetworkComputing2023}. In this context, several works have explored integrating KD to embed intelligence by compressing complex models directly into the data plane while respecting its stringent hardware and memory constraints. 

In \cite{linInNetworkFlowClassification2021}, the authors propose a two-phase learning framework that combines remote training with in-network inference via KD for flow classification. A teacher model is trained externally, while a lightweight student model is distilled and deployed on programmable switches assisted by Neural Compute Sticks (NCS). This architecture reduces inference rejection rates while preserving high accuracy. However, the approach depends on offline-trained teacher models, which may become outdated under dynamic traffic conditions.

In \cite{xieEmpoweringInNetworkClassification2024}, the authors introduce a teacher-student KD framework that transfers knowledge from high-capacity models, such as neural networks or ensemble learners, to a Binary Decision Tree (BDT). The distilled BDT is mapped onto ternary match tables, ensuring compatibility with the softwarized network hardware pipeline constraints while achieving improved classification accuracy and reduced memory footprint. However, this approach remains limited by the representational capacity of decision trees and the scalability constraints imposed by finite table entries.

The authors in \cite{demarinisCascadedLookTable2023} introduce a lossless KD technique that maps a trained Deep Neural Network (DNN) into a cascade of Look-Up Tables (LUTs) suitable for P4 pipelines. By transforming neuron computations into match-action operations, the approach eliminates arithmetic operations unsupported in hardware switches. Evaluated on a Distributed Denial of Service (DDoS) mitigation use case, the method demonstrates the feasibility of full DNN offloading. However, LUT size grows exponentially with input dimensionality and bit resolution, limiting scalability for high-dimensional feature spaces.

In \cite{liuGenerativeAiEnabledLightweight2025}, the authors present a GenAI-enabled architecture that leverages generative KD to train lightweight decision tree models for deployment on programmable wireless gateways. The framework aims to balance detection accuracy with data plane resource constraints for online traffic detection. While the approach demonstrates improved efficiency, its performance depends on the quality and representativeness of the generative model.

Across these works, teacher models are trained off-device while distilled students are deployed in the data plane under strict hardware constraints. KD thus acts not only as compression but also as a translation mechanism toward hardware-compatible representations such as decision trees or lookup tables. The main trade-off is between low-latency compatibility and model expressiveness, while reliance on offline teachers limits adaptability to dynamic traffic.

\subsubsection{Network slicing and service function intelligence}~\\
\indent KD has been increasingly adopted at the service layer of softwarized networks, particularly in network slicing, Service Function Chaining (SFC), resource allocation, and computation offloading.

In \cite{tangAnomalyDetectionService2024}, the authors propose a distributed KD  framework for anomaly detection in SFCs within industrial IoT environments. The proposed approach integrates distributed KD to improve SFC-level anomaly detection while reducing model redundancy. This improves detection accuracy while lowering training overhead through collaborative learning. It is, however, necessary to further investigate energy efficiency and heterogeneous deployment constraints.

In \cite{hossainLightweight5GV2XIntraSlice2023}, the authors propose a lightweight Intrusion Detection System (IDS) for 5G Vehicle-to-Everything (V2X) network slicing using KD. A large DL model is trained in the cloud and distilled into slice-specific lightweight models deployed at the edge. Experimental results show a reduction in computation and memory overhead while maintaining comparable detection accuracy. However, adaptability under dynamic configuration remains to be evaluated.

The authors in \cite{elormkuadeyDeepFedKDADDeepFederated2023} propose a deep federated KD framework for anomaly detection in Beyond 5G (B5G) network slicing. The framework combines complex DL models with federated KD to preserve privacy and reduce communication overhead. Results demonstrate improved accuracy, precision, recall, and faster convergence compared to baseline approaches. However, the KD component is not clearly detailed.

A traffic-aware hierarchical offloading framework for slicing-enabled Space-Air-Ground Integrated Networks (SAGINs) is proposed in \cite{chenTrafficAwareLightweightHierarchical2024}. DRL is integrated with KD to compress offloading policies into lightweight models suitable for resource-constrained nodes. Results show improved QoS metrics and reduced inference overhead compared to heavier DRL baselines.

In \cite{kwantwiPersonalizedFederatedLearning2025}, the authors investigate personalized FL for slice-based task offloading and resource allocation. The framework leverages KD to support lightweight personalized models across slices. Although the approach enhances adaptability and slice-specific performance, it also leads to an increase in aggregation complexity.

Across these works, KD compresses DL models and RL policies for multi-tenant and privacy-aware environments, with teachers typically hosted in centralized cloud environments and students deployed across slices or edge nodes. While KD reduces computational overhead and supports scalable deployment, it also increases coordination complexity. A key limitation is the lack of dynamic adaptation to traffic variability, mobility, and evolving service demands.

\subsubsection{Emerging AI-native programmable networks}~\\
\indent The evolution toward AI-native programmable networks has introduced large foundation models, generative AI, and agentic intelligence directly into softwarized network architectures. In this context, KD enables the practical deployment of large AI models within softwarized networks.

The authors in \cite{heAdvancingEndtoEndProgrammable2025} propose an end-to-end programmable network framework that integrates GenAI for network optimization and DDoS detection. They combine programmable data planes with generative models and apply KD to compress large models for deployment in resource-constrained environments. Simulation results demonstrate improvements in detection latency and accuracy. However, it is important to further investigate large-scale deployment overhead.

In \cite{liuLAMeTAIntentAwareAgentic2025}, the authors introduce an intent-aware agentic network optimization framework that distills large models into lightweight edge models using intent-oriented KD. The approach integrates distilled edge Large Language Models (LLMs) with DRL to translate user intents into structured preference vectors for network optimization. Experimental results show reduced intent-prediction error and improved Quality of Experience (QoE) over conventional DRL.

An LLM-empowered intent-driven network configuration generator integrating fine-tuning, Retrieval-Augmented Generation (RAG), prompt engineering, and KD is proposed in \cite{liLargeLanguageModelEmpowered2025}. The framework distills a lightweight model suitable for deployment on edge devices while maintaining configuration generation performance. Results show reduced generation time and improved accuracy compared to baselines.

In these architectures, KD translates large foundation models into lightweight edge-deployable counterparts. While this enables capabilities such as intent-aware optimization and semantic reasoning, it also raises challenges in energy consumption, inference latency, and edge deployment feasibility. Moreover, most works emphasize performance metrics, while energy efficiency remains underexplored.

\begin{table*}[t]
\centering
\caption{Comparative Analysis of Knowledge Distillation in Intelligent Softwarized Networks}
\label{tab:kd_comparison}
\resizebox{\textwidth}{!}{
\begin{tabular}{lllllll}
\toprule
\textbf{Year} & \textbf{Ref.} & \textbf{Target Function} & \textbf{KD Paradigm} & \textbf{Teacher Location} & \textbf{Student Location} & \textbf{Reported Improvement} \\
\midrule

2021 & \cite{linInNetworkFlowClassification2021} 
& In-network flow classification 
& Response-based 
& SDN controller 
& Switch (with NCS) 
& Reduced rejection rate \\

2023 & \cite{demarinisCascadedLookTable2023} 
& P4 DNN deployment (DDoS) 
& Response-based 
& Cloud/offline training 
& P4 switch 
& Hardware-feasible deployment \\

2023 & \cite{hossainLightweight5GV2XIntraSlice2023} 
& Slice IDS (5G-V2X) 
& Response-based 
& Cloud 
& Slice-level nodes 
& Reduced computational overhead \\

2023 & \cite{elormkuadeyDeepFedKDADDeepFederated2023} 
& Slice anomaly detection 
& Federated KD 
& Federated clients 
& Slice functions 
& Improved convergence and privacy preservation \\

2024 & \cite{xieEmpoweringInNetworkClassification2024} 
& In-network classification 
& Feature-based 
& Cloud/server 
& Programmable switch 
& Reduced memory footprint, improved accuracy \\

2024 & \cite{tangAnomalyDetectionService2024} 
& SFC anomaly detection 
& Distributed KD 
& Edge/cloud 
& SFC nodes 
& Maintained detection performance \\

2024 & \cite{chenTrafficAwareLightweightHierarchical2024} 
& Slice offloading (SAGIN) 
& Policy distillation 
& Centralized trainer (cloud) 
& Edge platforms 
& Reduced model complexity \\

2024 & \cite{salamiDistributedLearningWiFi2024} 
& Wi-Fi AP load prediction 
& Federated KD 
& Distributed APs 
& AP nodes 
& Reduced communication and energy overhead \\

2025 & \cite{liuGenerativeAiEnabledLightweight2025} 
& Traffic detection (gateway) 
& Generative KD 
& Cloud 
& Programmable gateway 
& Reduced lookup-table complexity \\

2025 & \cite{ameurExploringTeacherStudentLearning2025} 
& QoS routing (SDN) 
& Multi-teacher DRL distillation 
& Distributed agents 
& SDN controller 
& Improved routing performance \\

2025 & \cite{kwantwiPersonalizedFederatedLearning2025} 
& Slice resource allocation 
& Federated KD 
& Federated server 
& Slice nodes 
& Reduced latency and resource usage \\

2025 & \cite{liuLAMeTAIntentAwareAgentic2025} 
& Intent-aware optimization 
& LLM-based KD 
& Foundation model (cloud) 
& Edge LLM instance 
& Improved QoE \\

2025 & \cite{heAdvancingEndtoEndProgrammable2025} 
& GenAI-enabled network 
& Response-based 
& GenAI model (cloud) 
& Programmable infrastructure 
& Reduced detection latency \\

2025 & \cite{liLargeLanguageModelEmpowered2025} 
& Intent-driven configuration 
& LLM-based KD 
& LLM (cloud) 
& Network controller 
& Improved configuration accuracy \\

2025 & \cite{zhuAUVWirelessCluster2025} 
& Multi-agent SDN tracking 
& Multi-teacher DRL distillation 
& Centralized SDN controller 
& AUV agents 
& Improved tracking performance \\

2025 & \cite{chenSpatialTemporalTrafficPrediction2025} 
& Traffic prediction (SAGIN) 
& Continuous KD 
& SDN controller/cloud 
& Edge nodes 
& Reduced transmission overhead \\

\bottomrule
\end{tabular}
}
\end{table*}

\subsubsection{Distributed and decentralized network intelligence}~\\
\indent KD is increasingly used as a coordination and knowledge-transfer mechanism in distributed softwarized networks, enabling collaborative learning among controllers, agents, and edge nodes while supporting policy transfer and adaptation.

In \cite{zhuAUVWirelessCluster2025}, the authors propose a software-defined multi-teacher-student RL framework for multi-target tracking in Autonomous Underwater Vehicle (AUV) networks. The architecture integrates SDN with multi-agent RL. A reciprocal teacher-student mechanism accelerates convergence and improves resource utilization through multi-teacher distillation. However, it is necessary to further investigate the computational overhead of the proposed solution.

The authors in \cite{chenSpatialTemporalTrafficPrediction2025} address spatial-temporal traffic prediction in SAGINs using an adaptive KD scheme. The proposed framework supports continual adaptation under dynamic traffic patterns while reducing data transmission overhead in SDN-based architectures. Deployment latency and large-scale scalability remain to be further evaluated.

In \cite{ameurExploringTeacherStudentLearning2025}, the authors propose a distributed teacher-student framework for QoS routing in SDN, combining multi-agent DRL with KD. In the proposed approach, multiple teacher agents guide student agents, enabling efficient policy learning for dynamic routing decisions. Experimental results show significant improvements in delay and throughput compared to standalone multi-agent DRL. The work does not explicitly quantify computational or energy savings.

In \cite{salamiDistributedLearningWiFi2024}, the authors investigate distributed learning for Wi-Fi Access Point (AP) load prediction, comparing FL and KD. KD is employed as a distributed training paradigm, enabling clients to distill knowledge into a global model without sharing raw data. The results demonstrate competitive predictive accuracy while reducing overhead and energy consumption.

Across these works, KD enables distributed intelligence without centralizing data. The main trade-off is between collaboration efficiency and communication overhead, since KD still requires synchronization and coordination. Most approaches also assume stable communication, which may not hold in dynamic real-world environments.

\subsection{Comparative analysis and discussion}
Table \ref{tab:kd_comparison} provides a cross-layer comparison of existing KD integrations in intelligent softwarized networks, highlighting target functions, distillation paradigms, architectural placement, and reported efficiency gains. This reveals how KD evolves across layers and exposes trends and challenges.

Across layers, distillation paradigms follow a clear progression. In the data plane, KD is mainly response-based or feature-based, compressing complex models into hardware-feasible representations for programmable switches and gateways \cite{linInNetworkFlowClassification2021, demarinisCascadedLookTable2023, xieEmpoweringInNetworkClassification2024, liuGenerativeAiEnabledLightweight2025}. At the service layer, federated and distributed KD dominate, supporting slice-level optimization, SFC management, and anomaly detection \cite{tangAnomalyDetectionService2024, hossainLightweight5GV2XIntraSlice2023, elormkuadeyDeepFedKDADDeepFederated2023, kwantwiPersonalizedFederatedLearning2025}. AI-native networks extend KD to foundation-model compression, where LLMs or generative models distill semantic and intent-aware capabilities into lightweight models \cite{heAdvancingEndtoEndProgrammable2025, liuLAMeTAIntentAwareAgentic2025, liLargeLanguageModelEmpowered2025}. In distributed and control-plane intelligence, KD enables collaborative and multi-agent learning, accelerating RL convergence and reducing communication overhead \cite{ameurExploringTeacherStudentLearning2025, zhuAUVWirelessCluster2025, salamiDistributedLearningWiFi2024, chenSpatialTemporalTrafficPrediction2025}. Consistently, high-capacity teachers remain centralized, while lightweight students are deployed closer to the edge.

Evaluation practices vary significantly across these works. Data plane studies emphasize accuracy and hardware feasibility \cite{linInNetworkFlowClassification2021, demarinisCascadedLookTable2023, xieEmpoweringInNetworkClassification2024}; service-layer approaches introduce QoS (e.g., delay, throughput) and detection metrics \cite{hossainLightweight5GV2XIntraSlice2023, tangAnomalyDetectionService2024, chenTrafficAwareLightweightHierarchical2024}; AI-native systems consider QoE and intent prediction quality \cite{liuLAMeTAIntentAwareAgentic2025, liLargeLanguageModelEmpowered2025}; and distributed schemes assess communication and energy overhead \cite{salamiDistributedLearningWiFi2024, elormkuadeyDeepFedKDADDeepFederated2023}. However, the lack of unified evaluation frameworks prevents rigorous comparison across approaches.

Overall, the analysis highlights that effective integration of KD in softwarized networks requires a holistic perspective that jointly considers model design, deployment constraints, and system-level trade-offs.
\section{Open Challenges and Future Directions}
Although KD enables efficient intelligence in softwarized networks, several challenges remain before it can serve as a fully integrated architectural component. Existing works focus on isolated layers, with limited cross-layer integration and insufficient analysis of energy, scalability, robustness, and lifecycle aspects. These limitations indicate that KD is still primarily used as a model compression tool rather than a system-level mechanism for distributed network intelligence.

\textbf{Energy-aware distillation in edge environments:}
Energy efficiency remains a key concern, particularly in edge and IoT environments with strict power constraints. While many studies reduce model size or communication overhead, few analyze trade-offs between energy consumption, inference latency, and accuracy. These trade-offs depend on deployment decisions, such as teacher-student placement across cloud and edge nodes. Future work should explore hardware-aware and energy-adaptive KD mechanisms.

\textbf{Dynamic and adaptive distillation:}
Softwarized networks operate under dynamic conditions, including traffic variability and mobility. Most KD approaches assume static teacher-student relationships and offline training, limiting real-world applicability. This calls for online, continual, and context-aware distillation mechanisms.

\textbf{Cross-layer integration and orchestration:}
Current works apply KD within isolated layers. However, intelligent softwarized networks require coordinated intelligence across layers, raising challenges in orchestration and resource allocation. Future research should explore cross-layer distillation frameworks.

\textbf{Communication overhead and scalability:}
In distributed settings, KD reduces raw data exchange but still requires sharing distilled knowledge. This introduces communication overhead and synchronization costs that may limit scalability. This remains an open challenge.

\textbf{Security, trust, and robustness:}
KD is vulnerable to threats such as poisoned teachers, adversarial transfer, and privacy leakage. In decentralized environments, a lack of trust further complicates secure knowledge sharing. Future work should develop trust-aware and privacy-preserving KD frameworks.

Overall, addressing these challenges is essential to position KD as a scalable, robust, and energy-efficient enabler of intelligence in next-generation softwarized networks.

\section{Conclusion}
The convergence of network softwarization and ML is driving the evolution toward adaptive communication infrastructures, where KD plays a central role in enabling efficient intelligence. In this paper, we showed that KD is evolving beyond model compression into a system-level enabler for intelligent softwarized networks under strict latency, energy, and resource constraints. However, several key challenges remain before KD can be fully established as a foundational building block for next-generation intelligent softwarized networks.

As future work, we consider designing an energy-aware KD framework for constrained environments that enables continual learning, cross-layer orchestration, and distributed deployment across cloud, edge, and resource-constrained devices.

\bibliographystyle{IEEEtran}
\bibliography{IEEEbib}

\end{document}